\documentclass{article}
\usepackage{spconf,amsmath,graphicx,multirow,float}
\usepackage{booktabs}

\title{MOS-FAD: IMPROVING FAKE AUDIO DETECTION VIA AUTOMATIC MEAN OPINION SCORE PREDICTION}
\name{Wangjin Zhou$^{1}$, Zhengdong Yang$^{1}$, Chenhui Chu$^{1}$, Sheng Li$^2$, Raj Dabre$^2$, Yi Zhao$^{3}$, Tatsuya Kawahara$^{1}$}
\address{$^1$Graduate School of Informatics, Kyoto University, Sakyo-ku, Kyoto, Japan\\
  $^2$National Institute of Information and Communications Technology (NICT), Kyoto, Japan\\
  $^3$Kuaishou Technology, Beijing, China}
\begin{document}
\ninept
\maketitle
\begin{abstract}
Automatic Mean Opinion Score (MOS) prediction is employed to evaluate the quality of synthetic speech. This study extends the application of predicted MOS to the task of Fake Audio Detection (FAD) as we expect that MOS can be used to assess how close synthesized speech is to the natural human voice. We propose MOS-FAD, where MOS can be leveraged at two key points in FAD: training data selection and model fusion. In training data selection, we demonstrate that MOS enables effective filtering of samples from unbalanced datasets. In the model fusion, our results demonstrate that incorporating MOS as a gating mechanism in FAD model fusion enhances overall performance.

\end{abstract}
\section{Introduction}

Recent developments in text-to-speech (TTS)~\cite{wang2017tacotron, ren2019fastspeech, ren2020fastspeech, yu2019durian} and voice conversion (VC)~\cite{sisman2020overview} have made it feasible to create a human-like speech, which could be misused for malevolent purposes such as spoofing attacks. 
As a countermeasure, there is a growing emphasis on fake audio detection (FAD), which aims to distinguish fake audio from real audio \cite{yi2021half, ma2021continual, yamagishi2021asvspoof, todisco2019asvspoof, kinnunen2017asvspoof, yi2022add}.

Efforts for FAD tasks have focused on enhancing the acoustic front end to improve the effectiveness of FAD systems. Research has shown that well-designed acoustic features can effectively distinguish fake audio from real audio \cite{todisco2017constant,sahidullah2015comparison,xie2021siamese, martin2022vicomtech}. Concurrently, there are also studies concentrated on designing effective classification models to distinguish between real and fake audios \cite{lai2019assert,alzantot2019deep,lavrentyeva2017audio,he2016deep,liu2018darts,ge2021partially,baevski2020wav2vec,xie2021siamese,martin2022vicomtech}. 

%
Unlike the earlier approach, our core idea is to improve fake audio detection by leveraging external knowledge. 
MOS (Mean Opinion Score) is used to assess the quality of synthesized speech, so MOS will be a promising tool for detecting fake audio. Moreover, compared to the binary classification of fake audio detection, MOS offers a finer-grained evaluation criterion that can yield additional information. In this paper, we adopt the self-supervised learning (SSL) models \cite{SSL2018}, widely studied and performed well in various speech classification tasks \cite{superb}, as the foundational architecture for constructing the FAD classifier. To ensure the robustness of our final results and reduce dependence on the performance of a single SSL model, we conduct experiments with various SSL models. 
Building upon this SSL-based architecture, we explore various methods for integrating MOS scores into FAD tasks. This investigation ultimately leads to developing our innovative MOS-FAD framework, where MOS scores are used as gating mechanisms in model construction. This framework has achieved state-of-the-art performance.

We also extend the use of MOS scores to guide training data selection. This approach addresses two key challenges: Firstly, it alleviates the long-tail issue associated with imbalanced training data. Secondly, by utilizing MOS scores for data selection, the model can be trained on samples that pose a greater challenge in distinguishing between real and fake audio, ultimately enhancing the model's classification capabilities.

\begin{figure*}[t]
    \centering
	\includegraphics[scale=0.35]{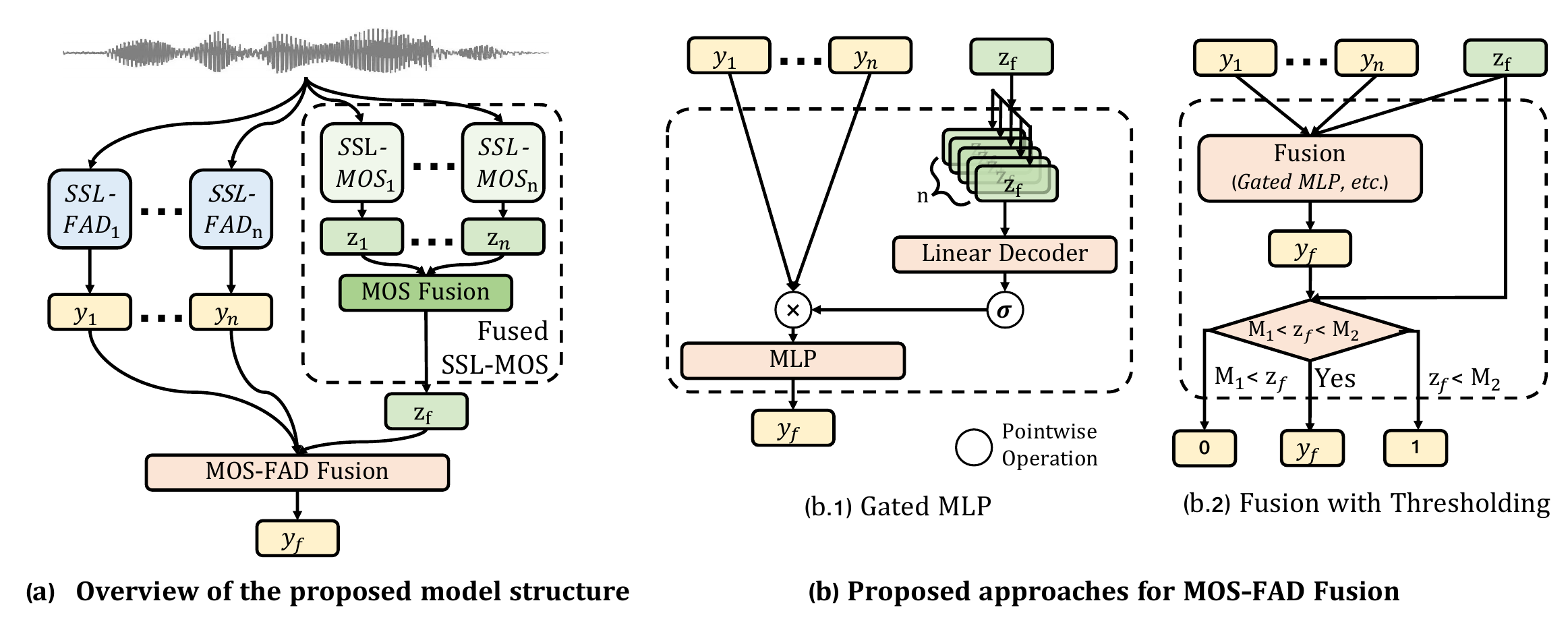}
	\caption{Proposed model structure.}
	\label{fig:model_proposed}
\end{figure*}

\section{Related work}
\subsection{Automatic MOS Predictor}

The Mean Opinion Score (MOS) is a widely used subjective quality evaluation criterion for synthesized speech. In MOS assessments, listeners rate speech sample quality on a scale ranging from 1 (poor) to 5 (excellent). The average of these ratings is used to gauge overall speech quality. However, traditional MOS evaluations can be expensive and time-consuming.

Given these challenges, there has been a growing interest in training automatic MOS predictors. Notably, the VoiceMOS Challenge \cite{huang2022voicemos} has advanced this field recently.

The state-of-the-art automatic MOS predictors employ neural network-based frameworks and use large-scale synthetic speech data for training a model to predict MOS score~\cite{lo2019mosnet, choi2021neural, huang2021ldnet, cooper2021generalization, Yang2022FusionOS, Saeki2022UTMOSUS}. Another solution is to introduce self-supervised learning (SSL) models~\cite{baevski2020wav2vec,chen2021wavlm,hsu2021hubert,chen2021wavlm,baevski2022data2vec}, which have already been demonstrated in the MOS prediction task. 
Since the choice of data and various model configurations used for pre-training will affect the performance, fusing diverse models with different architectures or training strategies \cite{Saeki2022UTMOSUS,Yang2022FusionOS} could help achieve better results on the MOS prediction task.

\subsection{Fake Audio Detection}
The current FAD research focuses on acoustic front-end and classification models. 

Research has shown that well-designed acoustic features can effectively distinguish fake audio from real audio. 
The traditional Mel frequency spectrum coefficients (MFCC) were improved by using constant Q-transform instead of the short-time Fourier transform \cite{todisco2017constant}. The linear frequency spectrum coefficients (LFCC) \cite{sahidullah2015comparison} was proposed to replace the Mel-scale filters with linear filters, making it more focused on high-frequency band features. Wav2vec pre-trained models \cite{xie2021siamese, martin2022vicomtech} were used as the feature extractors, which obtain more robust acoustic features.

An effective classification model can also help distinguish between real and fake audio. With the development of deep neural networks, the classifier has been upgraded from convolutional neural networks (CNN) \cite{lai2019assert, alzantot2019deep}, light convolution neural network (LCNN) \cite{lavrentyeva2017audio}, residual network (ResNet) \cite{he2016deep}, differentiable architecture search (DARTS) \cite{liu2018darts, ge2021partially} to state-of-the-art Transformer-based models, e.g., Wav2vec2.0 \cite{baevski2020wav2vec,xie2021siamese, martin2022vicomtech}.

\section{Proposed Approach}
As illustrated in Fig. \ref{fig:model_proposed}. (a), our proposed method consists of three components: SSL-FADs, Fused SSL-MOS sub-system, and MOS-FAD Fusion. 
In this section, we introduce each of them.

\subsection{SSL-FAD}

SSL-FADs are SSL-based models that predict FAD scores. This work adopts the idea in \cite{cooper2021generalization}, which adds a mean pooling layer and a fully connected layer after the feature extractor of an SSL model to construct an SSL-FAD model. 
In this paper, we adopt $7$ different SSL models (Wav2Vec 2.0 Base, Wav2Vec 2.0 Large, Wav2Vec 2.0 (LV-60), HuBERT Base, WavLM Base, WavLM Base+ and WavLM Large) to construct $7$ SSL-FADs, as shown in Table \ref{tab:ssl-fads}. 
By $n$ SSL-FADs, FAD scores $\mathcal{Y} = \{ y_i \mid y_i \in [0, 1], \, i = 1, 2, \ldots, n \}$ are computed.

\subsection{Fused SSL-MOS}
\label{subsec:ssl-mos}

The fused SSL-MOS is an automatic MOS predictor proposed in \cite{Yang2022FusionOS}, which consists of several SSL-MOS models and an MOS fusion. 
Like SSL-FAD, SSL-MOS is constructed by adding a mean pooling layer and a fully connected layer after the feature extractor of the SSL model. The SSL models employed in constructing SSL-MOSs are identical to those used in constructing SSL-FAD.
The MOS model fuser is a $2$-layer neural network consisting of a fully connected layer
without a bias for capturing the weighted information, and a linear function for obtaining the residual information between the ground truth and the predicted scores after a fully connected layer.  
The MOS fusion takes the MOS scores $\mathcal{Z} = \{ z_i \mid z_i \in [0, 5], \, i = 1, 2, \ldots, n \}$ predicted by $n$ SSL-MOS models as inputs and outputs the final MOS score $z_f \in [0, 5]$.

\subsection{MOS-FAD Fusion}
\label{subsec:mos-fad}

The MOS-FAD fusion performs the fusion of FAD scores predicted by the SSL-FADs and the MOS scores predicted by the Fused SSL-MOS. It outputs the FAD task's prediction $y_f \in [0, 1]$.
 
 We conducted experiments using both Multi-Layer Perceptron (MLP) and LightGBM~\cite{ke2017lightgbm} as potential candidates for MOS-FAD. Through a comparative analysis of their performance, we concluded that MOS can be effectively employed as a gating mechanism within the fusion model. Building upon this insight, we introduced Gated MLP, as depicted in Fig.\ref{fig:model_proposed}(b.1). Moreover, to further leverage the prior information from MOS scores, the Fusion with thresholding is proposed, as shown in Fig.\ref{fig:model_proposed}(b.2).

\noindent\textbf{a. MLP}: In our experiment, the MLP is a two-layer neural network with one sigmoid activation function and three hidden neurons. The linear transformation in the first layer is performed without a bias.

\noindent\textbf{b. LightGBM}: Light Gradient Boosting Machine (LightGBM)~\cite{ke2017lightgbm} is an ensemble model that trains a series of decision trees sequentially in a leaf-wise fashion and combines them. During sequential training, the decision trees use the error from the previous tree to adjust their learning and eventually minimize the loss function. Therefore, weak learners of the decision trees can be combined as a high-performance model. 

\noindent\textbf{c. Gated MLP}: Instead of feeding all the predicted scores (FAD scores and MOS scores) to the MLP, we design a special gate layer for the predicted MOS score. 
The gates control how much information from the FAD scores should be fed to the later MLP. 
The first step is to decode the MOS scores into a tensor of the same length as the FAD scores using a Linear Decoder layer. Then, the values in this tensor are normalized to the range of 0 to 1 using a $Sigmoid$ activation function. These outputs are treated as gate $g$. 
Then, predicted FAD scores $y_1$, $y_2$ ... are pointwisely multiplied by $g$, and the results are used as inputs to the MLP. The MLP component in Gated MLP remains consistent with the description provided in \ref{subsec:mos-fad}.a.

\noindent\textbf{d. Fusion with Thresholding}: Building upon the definition of MOS scores, we introduce a straightforward screening mechanism named as `fusion with thresholding'. The underlying concept is to categorize samples with excessively low MOS scores as fake audio and those with excessively high MOS scores as genuine audio. We assign a score of $0$ to samples with MOS scores below the threshold $M_1$, designating them as fake speech, and assign a score of $1$ to samples with MOS scores above the threshold $M_2$ as real speech. FAD scores are determined for the remaining samples based on the outcomes of specific fusion models. It is worth highlighting that this fusion with thresholding approach applies to any FAD model. In our experiments, all fake speech of training and validation datasets received MOS scores below 4.0, while 98.81\% of real speech received MOS scores above 2.5. Consequently, we set the threshold values as $M_1 = 2.5$ and $M_2 = 4.0$.

\begin{figure}[!t]
    \centering
	\includegraphics[scale=0.36]{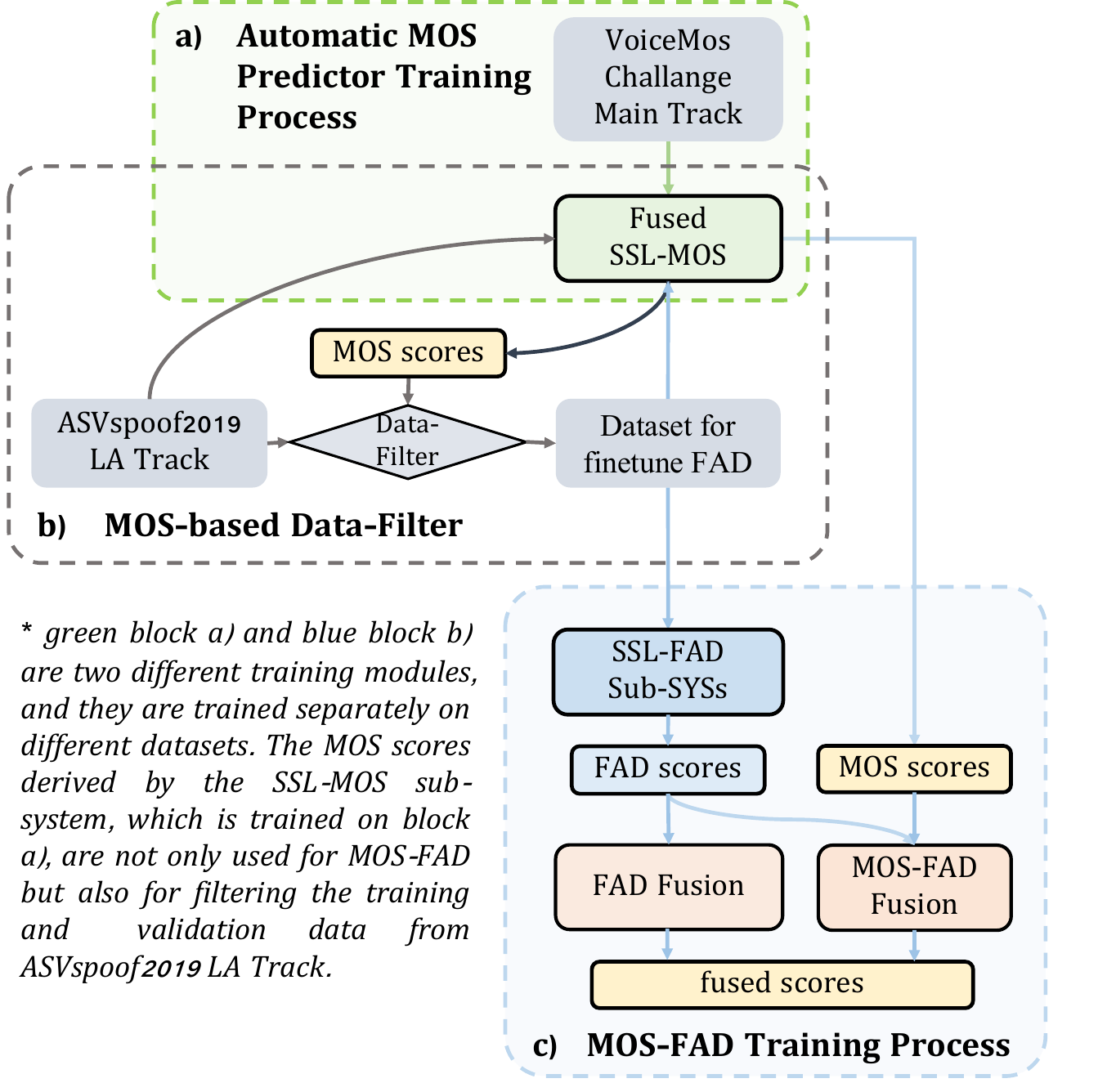}
	\caption{Training process of our proposed model.
 }
        \label{fig:implementation_details}
\end{figure}

\section{Experiment}
\label{sec:exp-con}

\subsection{Datasets}

\noindent\textbf{a. VoiceMOS dataset}: We used the VoiceMOS Challenge 2022 \cite{huang2022voicemos} main track dataset for training automatic MOS predictor.
The main track data contains only English and consists of $4,974$ samples for training, $1,066$ samples for validation, and $1,066$ examples for testing. 
We quantized the main track's MOS scores ranging from $1$ to $5$ into $33$ intervals with a step of $0.125$.

\noindent\textbf{b. ASVspoof dataset}: ASVspoof datasets were leveraged for the FAD task. 
First, we trained our proposed framework by using the training and validation datasets from the ASVspoof2019 \cite{todisco2019asvspoof} LA dataset. 

Next, we employed the ASVSpoof2021 challenge \cite{yamagishi2021asvspoof} evaluation set of the DF track to evaluate the effectiveness of our approach. This DF evaluation set comprises a total of $611,829$ samples. It shows audio coding and compression artifacts, with approximately $600$k of audio processing with various commercial audio codecs.

\subsection{Experimental Settings}

All neural network models were trained using a fixed learning rate of $0.001$, Stochastic Gradient Descent as the optimizer, and $CrossEntropy$ as the loss function. Training would stop if there were no decrease in the validation loss for 20 consecutive epochs. During the training of LightGBM, the following parameter configuration was employed: \{`objective': `binary', `metric': `auc', `num\_leaves': $16$, `max\_bin': $25$, `max\_depth': $4$, `learning\_rate': $0.1$\}.

\subsection{Training Process}
\label{subsec:train_proecess}

Fig.~\ref{fig:implementation_details} shows the training details of our proposed approach. 
The training process can be divided into two independent modules, the Automatic MOS Predictor Training Process and the MOS-FAD Training Process, and an additional data processing module, the MOS-based data-filter, for the FAD task training dataset. 

\noindent\textbf{a. Automatic MOS Predictor Training Process}: 
As depicted in the green block in Fig.~\ref{fig:implementation_details}, we employed the MOS fusion model, as described in Section \ref{subsec:ssl-mos}, to generate MOS scores. This model undergoes training using the VoiceMos Challenge 2022 Main Track dataset. It is important to note that this training process is conducted separately from the MOS-FAD training process, and the training specifics for SSL-MOS closely adhere to those detailed in \cite{Yang2022FusionOS}.

\begin{figure}[t]
    \vspace{-0.2cm}
    \centering
    \includegraphics[scale=0.3]{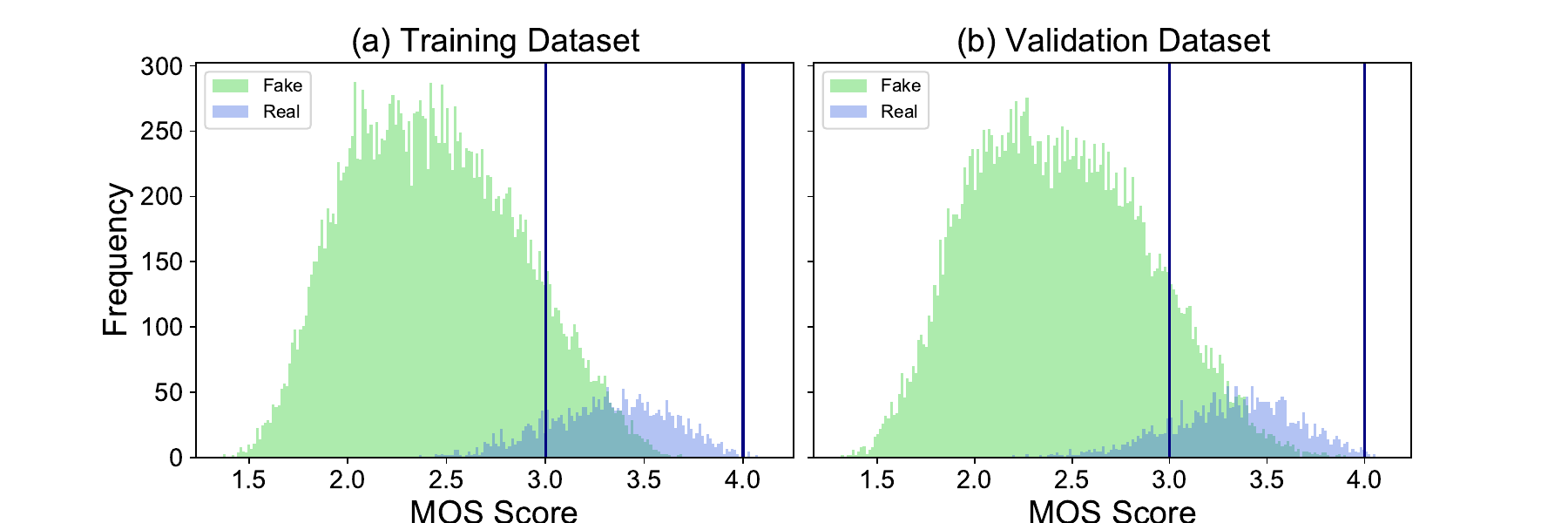} 
	\caption{The MOS score distributions in ASVspoof2019 LA track.}
	\label{fig:data}
\end{figure}

\begin{table}[t]
\setlength\tabcolsep{4.2pt}
\centering
\caption{The number of samples of ASVspoof2019 LA Track with or without the data-filter. With the data-filter we select data by choosing data with MOS scores ranging between $3.0$ and $4.0$. }
\begin{tabular}{c|rrr|rrr}

\hline
 & \multicolumn{3}{c}{\textbf{Training Dataset}} & \multicolumn{3}{|c}{\textbf{Validation Dataset}} \\
 & Total & Real & Fake & Total & Real & Fake \\
\hline
w/o filter & 25,380 & 2,580 & 22,800 & 22,438 & 2,548 & 22,438 \\
w/ filter & 5,034 & 2,568 & 2,466 & 5,198 & 2,533 & 2665 \\
\hline

\end{tabular}
\label{tab:resmaple_result}
\end{table}

\noindent\textbf{b. MOS-based Data-Filter}: The MOS-based data-filter serves as a selector for training FAD task models, with two primary objectives. Firstly, it aims to alleviate the issue of sample imbalance, as evident in Table~\ref{tab:resmaple_result}, where the original training set exhibits a significant imbalance, with nearly ten times more fake speeches than real ones. Secondly, it enhances model training effectiveness, as illustrated in Fig.~\ref{fig:data}. The distribution of MOS scores in the LA track dataset suggests that samples with MOS scores below $2.5$ can be confidently categorized as fake, while those with scores exceeding $4.0$ can be considered as real. Consequently, regarding its authenticity, the FAD model only requires training on data that falls within the range of uncertainty (overlap part in the MOS distribution graph).

In summary, to achieve a balanced sample distribution and enhance training effectiveness, we opted for samples from the ASVspoof2019 LA dataset with MOS scores between 3.0 and 4.0 as the new training and validation datasets for the FAD task. As depicted in Table~\ref{tab:resmaple_result}, this filtering approach results in an approximate balance between real and fake speeches.

\noindent\textbf{c. MOS-FAD Training Process}: We used the filtered dataset to finetune SSL-FADs and train their fusions (FAD Fusion and MOS-FAD Fusion) for FAD tasks (as shown in Fig.~\ref{fig:implementation_details}).  
The MOS-FAD Fusion in Fig.~\ref{fig:implementation_details} represents a model fusion approach that combines MOS and FAD scores. Concurrently, The FAD Fusion in Fig.~\ref{fig:implementation_details} serves as a control group, enabling us to evaluate the importance of incorporating MOS scores into the fusion model.

\subsection{Inference}

To assess the authenticity of samples in the ASVspoof2021 DF evaluation set, we first employ the Fused SSL-MOS system to generate its MOS scores. Concurrently, we use individual SSL-FAD systems to derive the corresponding FAD scores. Subsequently, the MOS-FAD Fusion process combines these FAD and MOS scores to yield a unified score for each sample. Notably, a lower fused score indicates a higher likelihood that the sample is fake.

\begin{table}[t]
\setlength\tabcolsep{4.2pt}
\centering
\caption{EER on 7 FAD scores of ASVSpoof2021 DF Eval Track by each SSL-FAD which trained with or without MOS Data-Filter. }
\begin{tabular}{crr}
\hline
 & \multicolumn{2}{c}{DF Eval} \\
 \textbf{SSL Base} &  \textbf{w/ filter} & \textbf{w/o filter} \\
\hline
 W2V Base  & 0.2514 & 0.5510\\
 HuBERT Base & 0.2353 & 0.4449\\ 
 W2V large & \textbf{0.1891} & 0.4794\\ 
 W2V Large(LV-6.0) & 0.2395 & \textbf{0.3776}\\ 
 WavLM Base & 0.2167 & 0.5126\\ 
 WavLM Base+ & 0.2088 & 0.5310\\ 
 WavLM Large  & 0.2453 & 0.5881\\ \hline
\end{tabular}
\label{tab:ssl-fads}
\end{table}

\section{Results and Analysis}
\label{sec:exp-result}

\subsection{Effectiveness of MOS-based Data-Filter}
\label{subsec:result-filter}

We conducted experiments by training SSL-FADs directly using the original training dataset and compared their performance with models trained on the filtered dataset, as displayed in Table~\ref{tab:ssl-fads}. The results demonstrate that models trained after data filtering outperform those trained on the unfiltered dataset, indicating the effectiveness of filtered data in training.

\subsection{Effect of MOS in Fusion}

Table~\ref{tab:results1} shows the performance of different fusion models.
Among them, the `Score Fusion' investigates how to combine $7$ different FAD scores and $1$ of MOS scores, while the `Embedding Fusion' validates the effectiveness of `Score Fusion'.

\noindent\textbf{a. Score Fusion}: Only using FAD scores, both FAD Fusion methods fail to outperform the Wav2Vec Large-based SSL-FAD model. 
This could be attributed to the high correlations among FAD scores, which indicate multicollinearity, making it challenging to improve performance through score fusion.

In contrast, within the MOS-FAD Fusions that incorporate MOS scores, LightGBM outperforms MLP and any SSL-FADs. Considering LightGBM's tree-like structure, we reason that MOS plays a gating role in the fusion process. Building on this hypothesis, we introduce the Gated MLP, which outperforms LightGBM, providing evidence of MOS as a gating mechanism.

For further improvement, we introduced a more rigorous gate filtering mechanism named `fusion with thresholding,' drawing upon prior MOS knowledge. Among MOS-FAD Fusions, the adoption of fusion with thresholding yields performance improvements. Notably, both LightGBM with thresholding and Gated MLP with thresholding surpass the previous SOTA performance. Gated MLP with thresholding demonstrates an impressive $13.6\%$ reduction in EER. 

\noindent\textbf{b. Embedding Fusion}: 
We substituted the MOS and FAD scores with embeddings generated by the SSL model, except for the MOS $z_f$ in `fusion with thresholding.' In the MLP approach of FAD Fusion, all embeddings from SSL-FADs are concatenated into a 6,144-dimension tensor, which is then input into the MLP. In the MLP approach of MOS-FAD Fusion, embeddings from both SSL-FAD and SSL-MOS are concatenated to create a 12,288-dimension tensor as input for the MLP. In the case of Gated MLP, the concatenated 6,144-dimension embeddings generated by SSL-MOS undergo a linear layer followed by a Sigmoid activation layer to produce gate units of 6,144 dimensions. 
These gate units control how much information from the concatenated 6,144-dimensional embeddings produced by SSL-FADs is passed into the MLP.
Notably, the number of hidden units in the MLP described in this section is half of the input dimension.

The results show that the embeddings-based MLP without MOS data failed to surpass the performance of that with MOS data and the best SSL-FAD model, highlighting the necessity of incorporating MOS information. 
Furthermore, embeddings-based fusions with thresholding outperform their respective baseline models, further validating MOS's appropriateness as the gating mechanism in fusion. 

Finally, when the significance level $p$ is set to $0.01$, the best performance of `Score Fusion' significantly outperforms that of `Embeddings Fusion.' A possible reason is that the gate logic in score fusion serves as a more powerful mechanism for combinations than the combinations learned through the model in embedding fusion. This confirms the effectiveness of `Score Fusion.'

\begin{table}[t]
\setlength\tabcolsep{4.2pt}
\centering
\caption{EER of each fusion on ASVSpoof2021 DF Eval Track. Scores stand for the fusion of FAD and MOS scores. Emb stands for the fusion of FAD and MOS embeddings.  $\dag$ is significantly better than the former SOTA when significance level $p$ is set at 0.01.
}
\begin{tabular}{ccrc}
\hline
 & & \multicolumn{2}{c}{DF Eval} \\
  \textbf{Model Type}  & \textbf{Model} & \textbf{Score} & \textbf{Emb} \\
\hline
\multirow{2}{*}{FAD Fusion} & MLP & 0.2110 & 0.2049\\
                         & LightGBM & 0.2221 & \textbackslash \\ \hline
\multirow{6}{*}{\shortstack{MOS-FAD\\Fusion}} & MLP &  0.3160 & 0.1763\\
                         & LightGBM & 0.1732 & \textbackslash \\
                         & Gated MLP & 0.1661 & 0.1890\\
                         & MLP + threshold & 0.2921 & 0.1509$^\dag$ \\
                         & LightGBM + threshold & 0.1476$^\dag$ & \textbackslash \\
                         & Gated MLP + threshold & \textbf{0.1351}$^\dag$ & 0.1509$^\dag$ \\
                         \hline
\multicolumn{2}{l}{Former SOTA in ASVspoof 2021 (T23 in \cite{yamagishi2021asvspoof})}  & \multicolumn{2}{c}{0.1564} \\ \hline
\end{tabular}
\label{tab:results1}
\end{table}
\vspace{0pt}

\section{Conclusion}
\label{sec:con}
In this paper, we have proposed MOS-FAD, a robust approach for FAD via automatic MOS predictor. We used the MOS-based data-filter to resample the training and validation dataset, which led to effective training for SSL-FADs. 
Additionally, our results have shown that utilizing MOS as a gating mechanism in the fusion with FAD scores improves performance, based on which we proposed a novel fusion model, Gated MLP, with thresholding.
This MOS-FAD fusion method, significantly decreases the EER by $13.6\%$, compared with the best system reported so far.
Given that the misuse of fake audio generation can potentially negatively impact society, we hope this study will encourage further research into this rapidly evolving and crucial field and help circumvent any problems.

\newpage

\bibliographystyle{IEEEbib}\footnotesize
\bibliography{mybib,add,acmart}

\end{document}